%% file: ms.tex
\pdfoutput=1	%
\documentclass[9pt,twocolumn,twoside]{pnas-new}
\setboolean{displaywatermark}{false}

\templatetype{pnasresearcharticle} %

\title{Newton: A Language for Describing Physics}

\author[1]{Jonathan Lim}
\author[2]{Phillip Stanley-Marbell} 

\affil[1]{Yelp, San Francisco, CA, USA}
\affil[2]{Department of Engineering, University of Cambridge, Cambridge CB3 0FA, UK.}

\leadauthor{Stanley-Marbell} 

\authorcontributions{PSM and JL designed and implemented the Newton Language. JL implemented the host language compiler API and evaluated its performance. Both authors contributed to writing this article.}
\correspondingauthor{\textsuperscript{2}To whom correspondence should be addressed. E-mail: {phillip.stanley-marbell@eng.cam.ac.uk}}

\vspace{-0.1in}

\input{newton-preamble.tex}
\input{abstract.tex}

\dates{This manuscript was compiled on \today}

\begin{document}

\verticaladjustment{-5pt}

\maketitle
\thispagestyle{firststyle}
\ifthenelse{\boolean{shortarticle}}{\ifthenelse{\boolean{singlecolumn}}{\abscontentformatted}{\abscontent}}{}

\vspace{-0.1in}
\dropcap{S}{ensor} data in physical systems are constrained by the laws of physics.
\input{introduction.tex}
\input{background.tex}
\input{design.tex}

\input{summary.tex}
\acknow{This research was initiated when the authors were both with
the Computer Science and Artificial Intelligence Laboratory (CSAIL),
Massachusetts Institute of Technology. The authors thank Zhengyang
Gu for contributing an SMT-based analysis pass~\cite{ZguMengThesis},
Vlad Mihai Mandric, Youchao Wang, and James Rhodes for assisting
with the implementation of linear-algebraic dimensional analysis
passes, and Rae Zhao for beta-testing the syntax for Newton sensor
descriptions.  PSM is supported in part by an Alan Turing Institute
award TU/B/000096 under EPSRC grant EP/N510129/1 and by Royal
Society grant RG170136.}

\showacknow %

\vspace{0.25in}

\bibliography{newton}

\onecolumn
\appendix

\twocolumn

\end{document}

%% file: newton-preamble.tex
\usepackage{framed}
\usepackage{moreverb}
\usepackage[small,sf,bf]{subfigure}
\usepackage{xspace}
\usepackage{flushend}			%

\usepackage{xfrac}			%

\usepackage[ruled,vlined]{algorithm2e}

\usepackage{amsfonts}
\usepackage{amssymb}    
\usepackage{amsmath}
\usepackage[thmmarks,amsmath,hyperref]{ntheorem}
\usepackage{graphicx}
\usepackage{times}
\usepackage[scaled]{helvet} 		%
\usepackage{textcomp}
\usepackage{color}
\usepackage{booktabs}
\usepackage{listings}
\usepackage{multicol}
\usepackage{times}
\usepackage{pifont}			%
\usepackage{microtype}

\usepackage{mmm}
\usepackage{wand-sample}
\usepackage[scaled=0.85]{beramono}
\usepackage[T1]{fontenc}
\usepackage{url}
\usepackage{multirow}
\usepackage{array}

\newcommand\code[1]             {{\bf\ttfamily\small #1}}

\definecolor{listinggreen}{rgb}{0,0.6,0}
\definecolor{listinggray}{rgb}{0.5,0.5,0.5}
\definecolor{listingmauve}{rgb}{0.58,0,0.82}
\definecolor{listingkeywordcolor}{rgb}{1.0,0.4,0.0}
\definecolor{listinglightgray}{rgb}{0.9863,0.9863,0.9863}

\lstdefinelanguage{FSharp}
{
	morekeywords	= {
    let,
    type,
    Measure,
	},
	sensitive	= false,
	morecomment	= [l]{\#},
	morecomment	= [s]{(*}{*)},
}

\lstdefinelanguage{Newton}
{
	morekeywords	= {
		signal,
		derivation,
		symbol,
		name,
		invariant,
		constant,
		English,
		sensor,
		name,
		none,
		dot,
		cross,
		derivative,
		integral,
		interface,
		i2c,
		spi,
		analog,
		write,
		read,
		delay,
		range,
		erasuretoken,
		uncertainty,
		accuracy,
		precision,
		Gaussian,
		exponential,
		biexponential,
		to,
		bits,
		dimensionless,
		include
	},
	sensitive	= false,
	morecomment	= [l]{\#},
	morecomment	= [s]{/*}{*/},
}

\usepackage{hyperref}
\hypersetup{
    colorlinks=false, %
    linktoc=all,     %
    linkcolor=blue,  %
}

%% file: abstract.tex
\begin{abstract}
Data from multiple sensors within an embedded computing system are
often constrained in their relationship to each other. They are
typically also constrained by general physical laws and by the
structural design (e.g., size and aspect ratio) and materials
properties (e.g., Young's modulus) of the objects to which they are
attached.  Today, computing in embedded sensor platforms ignores
this potentially rich source of information.

\qquad One way to exploit information about the physical context
in which a given sensor-driven program will be deployed, is to
modify the program (or its compiler) to implement algorithmic
transformations that exploit information about inter- and
intra-sensor-signal relationships imposed by physical structure and
materials properties. Alternatively, information about the physics
of signals could be provided as a specification, in a machine-readable
form, to be used by compilers and other analysis tools.

\qquad This article introduces Newton, a specification language for
notating the analytic form, units of measure, and sensor signal
properties for physical-object-specific invariants and general
physical laws. We designed Newton to provide a means for hardware
designers (e.g., sensor integrated circuit manufacturers, computing
hardware architects, or mechanical engineers) to specify properties
of the physical environments in which embedded computing systems
will be deployed (e.g., a sensing platform deployed on a bridge
versus worn by a human). Compilers and other program analysis tools
for embedded systems can use a library interface to the Newton
compiler to obtain information about the sensors, sensor signals,
and inter-signal relationships imposed by the structure and materials
properties of a given physical system. The information encoded
within Newton specifications could enable new compile-time
transformations that exploit information about the physical world.
\end{abstract}

%% file: introduction.tex
\label{Introduction}
Sensor data are also constrained by the mechanical design and
materials properties of the objects in which sensors are embedded
or to which they are attached. As a result, algorithms that process
sensor data, such as the algorithms in pedometers, unmanned aerial
vehicles, autonomous land vehicles, robots, and more, consume input
data that are constrained by physics.
Compilers for embedded computing systems could exploit this observation
to simplify arithmetic operations or to improve reliability in
sensor signal processing.

Incorrect sensor readings affect systems built on top of those
readings and can have catastrophic consequences: almost half of the
accidents related to industrial chemical processes in one
study~\cite{france} were attributed to errors in temperature and
pressure sensor readings. Similarly, erroneous pressure sensor
readings have led directly to aviation accidents~\cite{PitotTubeAccident,
macpherson1998black}.

Information on physical constraints on a system's sensor data could
allow runtime assertions on the sensor data, analogous to probabilistic
assertions~\cite{PLDI-2014-SampsonPMMGC}. Information on physical
constraints could also allow compile-time transformations that
substitute code that reads from sensors of one type, with code that
reads from sensors of a another type. Such sensor substitution is
analogous to strength reduction in traditional compiler
optimization~\cite{Cocke:1977:ARO:359863.359888} and could have
significant benefits to the energy efficiency and cost of sensor-driven
systems.

\begin{figure}
\centering
\includegraphics[width=0.485\textwidth]{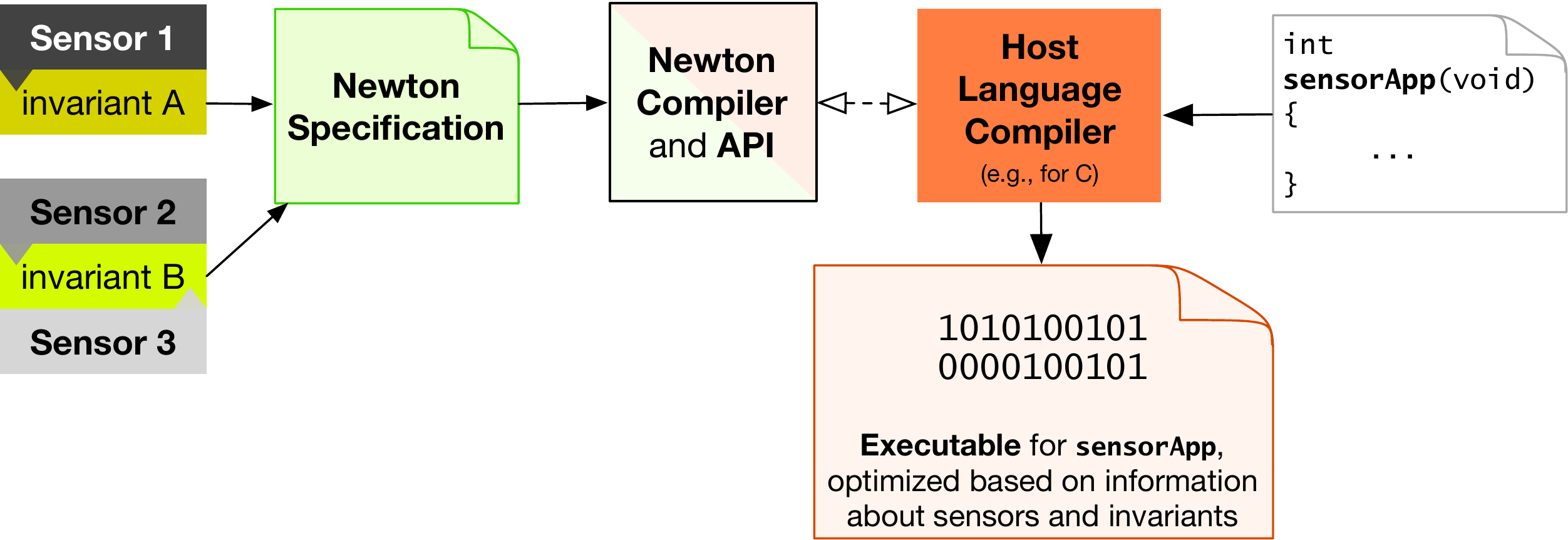}
\vspace{-0.15in}
\caption{The Newton compiler makes the intermediate representation
it generates from a Newton specification available to a \emph{host
language compiler} (e.g., a C compiler), through a library interface.
A host language compiler need not be changed to accommodate new
invariants in different sensor platforms.  A host language compiler
can also instrument the executables it generates, to make calls to
the Newton runtime library, in order to implement invariant checking
at runtime.}
\vspace{-0.1in}
\label{withNewton}
\end{figure}

To exploit these physical constraints that exist on sensor data, compilers
of embedded programming languages require specifications of those
contraints (Figure~\ref{withNewton}).

\section{Newton}
\vspace{-0.05in}
Newton is a language for specifying
dimensionally-annotated constraints and invariants on values
obtained from sensors embedded in physical structures. The Newton
specification compiler provides a library interface that programming
language compilers can use to obtain information about the physical
constraints on the signals in the programs they process. Compiled
programs and runtime systems can also access information about their
physical environments and sensor invariants using Newton's runtime
library, which provides routines for runtime querying of invariant
properties~\cite{JonathanLimMengThesis}.

\vspace{-0.05in}
\subsection{Example}
The example below shows the Newton specification for the
relationship between the period of an idealized simple pendulum,
the distance between the pivot point of the pendulum and its
primary mass, and constants (e.g., acceleration due to gravity).

\begin{lstlisting}[language=Newton]
time: signal  = {
   name       = "second" English;
   symbol     = s;
   derivation = none;
}

length: signal = {
   name        = "meter" English;
   symbol      = m;
   derivation  = none;
}

mass: signal  = {
   name       = "kilogram" English;
   symbol     = kg;	
   derivation = none;
}

Pi : constant = 3.14;
g  : constant = 9.8*m*s**-2;

pendulum: invariant(L: length, period: time) = {
   period ~ 2*Pi*((L/g)**(1/2))
}
\end{lstlisting}

%% file: background.tex
\section{Related Research}
\label{Background}
Dimensional analysis has been a valuable tool in science and
engineering disciplines for over a century. Early work by
Buckingham~\cite{PhysRev.4.345}, laid the foundations for the more
systematic study and application of dimensions in both design and
system evaluation. Central to many modern applications of dimensional
analysis has been the Buckingham Pi Theorem, which states that any
constraint between $n$ physical quantities, comprising $k$
dimensionally-independent physical quantities, can be represented
using a reduced number, $n - k$, of monomial expressions constructed
from the original $n$ physical quantities.

In programming languages, introducing dimension types has been
explored through both built-in types and libraries.
House~\cite{House:1983:PEF:4741.4742} proposed extending the Pascal
language with units and dimensions, and F\#~\cite{Kennedy_Fsharp}
includes mechanisms for a programmer to declare dimensions of
physical quantities. F\# allows programmers to use these dimension
types in programs, and provides support for type checking and type
inference based on a dimension unification
algorithm~\cite{Kennedy_DimensionTypes}.  Similarly,
XeLda~\cite{Antoniu:2004:VUC:998675.999448} provides type checking
for data in Excel spreadsheet programming.

Unlike techniques and systems for expressing and checking dimensions
in programming languages, the objective of Newton is instead to
allow sensor manufacturers, embedded system hardware platform
manufacturers, industrial design engineers, and physicists to
describe mechanical and other physical invariants obeyed by physical
artifacts instrumented with sensors. Programming language compilers
can then use these specifications to guide static compile-time and
dynamic run-time program transformations.  Newton enables compilers
to extend their use of physical information beyond dimension type
checking: Using information on sensor signal relationships obtained
from Newton at compile time or at runtime, code generated by compilers
that use Newton could check not only dimensions, but also physics-derived
and platform-specific signal property invariants.

%% file: design.tex
\section{Newton Descriptions}
\label{Design}
There are three components of a Newton description: signal
definitions, constant definitions, and invariant definitions.

\subsection{Signals}
Signals in Newton define either fundamental or derived signals, and
their units. They typically represent signals that can be read from
sensors on a hardware platform, such as acceleration from accelerometers,
magnetic flux density from magnetometers, and angular rate from
gyroscopes. Some fundamental signal types (e.g., \code{time}) exist
primarily to be able to define derived signal types relevant to
sensors, while others directly represent the signals of sensors
(e.g., \code{temperature}). The Newton language does not specify
which signals are fundamental \textit{a priori}, and a Newton
specification can define its own choice of fundamental signals. The
compiler installation provides a standard set of signal definitions
and in practice most Newton descriptions build on top of this
standard set of definitions.

All signal definitions in Newton have a \code{derivation} statement.
Fundamental signals have \code{none} as their \code{derivation},
while non-fundamental signals have a \code{derivation} which is a
monomial expression comprising previously-defined fundamental or
derived signals.

Newton specifications can define multi-dimensional signals, as the
examples of the signals \code{distance} and \code{speed} below
illustrate. Example uses of multi-dimensional signals include sensors
such as 3-axis accelerometers and specifying signals from a single
sensor sampled at different locations in space.

\vspace{1.0in}
\vfill

\begin{lstlisting}[language=Newton]
time : signal  = {
    name       = "second" English;
    symbol     = s;
    derivation = none;
}

distance : signal(i: 0 to 2) = {
    name       = "meter" English;
    symbol     = m;
    derivation = none;
}

speed : signal(i: 0 to 2) = {
    derivation = distance@i / time;
}
\end{lstlisting}

The \code{name} field of a signal definition is a human-readable
word or phrase describing the signal in a specific set of units and
therefore includes a language designator (e.g., \code{English}).
The \code{symbol} field in a signal definition specifies a token
that can be used as an alias for a specific unit for the signal.
For example, \code{distance} in the example above is defined as a
fundamental signal (\code{derivation = none}) with units description
\code{"meter"} and unit symbol \code{m}. This approach is similar
to aliasing in F\# \cite{Kennedy_Fsharp}.

\vspace{-0.05in}
\subsection{Constants}
Constants are fixed values with a dimension formed from a monomial expression
of previously-defined signals. Constants can also be
dimensionless, such as the mathematical constant $\pi$:

\begin{lstlisting}[language=Newton]
speedLimit	: constant = 100 * m / s;
Pi		: constant = 3.1415926535897932384626433832795;
\end{lstlisting}

\vspace{-0.15in}
\subsection{Invariants}
Invariant definitions take in a list of parameters (signals and
constants) with designated dimensions and define a physical relationship
between those parameters and previously-defined signals and constants.
The bodies of Newton invariants are comma-separated lists of
expressions involving a relational operator and these
expressions are interpreted as being in a conjunction.